%% file: lfox_MB-editedbyNS.tex
\documentclass[mypaper,7pt,twoside]{CoAst}
\usepackage{epsf,graphicx,fancyhdr}
\input{CoAst_regularlayo}

\begin{document}
\sf

\chapterCoAst{On the nature of HD 207331: a new $\delta$ Scuti variable}
{L.\,Fox Machado, W.J. Schuster, C. Zurita, et al.} 
\Authors{L.\,Fox Machado$^{1}$, W.\,J.\, Schuster$^1$,
C.\,Zurita$^2$,  J.\,L.\, Ochoa$^{1}$, and J.\,S.\,Silva$^{1}$}
\Address{$^1$ Observatorio Astron\'omico Nacional, Instituto de
Astronom\'{\i}a, Universidad Nacional Aut\'onoma de M\'exico,
Ensenada B.C., Apdo. Postal 877,
Mexico,\\
$^2$ Instituto de Astrof\'{\i}sica de Canarias, E-38205 La Laguna,
Tenerife, Spain\\}

\noindent
\begin{abstract}

While testing a Str\"omgren spectrophotometer attached to the 1.5-m
telescope at the San Pedro M\'artir observatory, Mexico, a number of A-type
stars were observed, one of which, HD 207331, presented clear
indications of photometric variability.  CCD  photometric data
acquired soon after
, confirmed its variability. In order to
determine its pulsation behaviour  more accurately,  $uvby$
differential photoelectric photometry was carried out for three
nights.  As a result of the period analysis of the light curves we
have found a dominant pulsation mode at 21.1 cd$^{-1}$ with an
amplitude of 6 mmag. This 
strongly suggests that HD 207331 is a new
$\delta$ Scuti-type pulsating star.
\end{abstract}

\Objects{HD 207331, BD$+$42 4208, TYC 3196-1243-1, HD 208310, HD
209113}

\section*{Introduction}
The $\delta$ Scuti-type pulsators  are stars with masses between 1.5
and 2.5 $M_{\odot}$ located at the intersection of the classical
Cepheid instability strip with the main sequence. These variables,
pulsating with radial and nonradial modes excited by the
$\kappa$ mechanism, are considered to be excellent laboratories for
probing the internal structure of intermediate mass stars. Thus, any
new detection of a $\delta$ Scuti star can be a valuable
contribution to asteroseismology.

\medskip
 HD 207331 ($=$ SAO 51294, BD$+$42 4207, HIP 107557) is  classified
in the SIMBAD database as a normal A0 star with apparent magnitude
$V$ of 8.31~ mag.  The Hipparcos catalogue (Perryman et al. 1997), on
the other hand, lists an $H_{p}$ of 8.3970 $\pm$ 0.0022 mag (median
error), $\pm$ 0.019 mag (scatter), a $V_{T }$ of 8.335 $\pm$ 0.009
mag (standard error), and a $(B-V)_{J}$ of 0.217 $\pm$ 0.011 mag
(standard error). Although it may present a possible variability given
that $H_{p, {\rm max}} = 8.37$ mag and $H_{p, {\rm min}} = 8.43$
mag reported in this catalogue, no period nor classification of
variability had been assigned to date.

\medskip
In the present paper, a study of the photometric variability of HD
207331 is reported.

\section{Observations and data reduction}
The observations were carried out at the Observatorio Astron\'omico
Nacional, San Pedro M\'artir,  Baja California, Mexico.
 The photometric variability
of HD 207331 was established on the night of 
September 27, 2007, using the six-channel $uvby-\beta$ spectrophotometer attached to the
H.L. Johnson 1.5-m telescope. The few data clearly 
show indications of photometric variability for this star.

\begin{table}[!t]\centering
  \setlength{\tabcolsep}{1.0\tabcolsep}
 \caption{Position, magnitude and spectral type of target, comparison and check
 stars observed in the CCD frame.}
  \begin{tabular}{llcccl}
\hline
Star&  $ID$ & RA & Dec & V  &  SpTyp \\
&&(2000.0)&(2000.0)&(mag)&\\
\hline
Target &HD 207331& 21 47 02   &$+$43 19 19 &8.3  & $A0$   \\
Comparison & BD$+$42 4208 & 21 47 12   &$+$43 19 51&9.4  &  $A0$   \\
 Check & TYC 3196-1243-1& 21 47 06    &$+$43 18 58  & 10.9 &- \\
\hline
\end{tabular}
\end{table}

\subsection*{CCD differential photometry}
CCD photometric observations of HD 207331 
confirming its
variability were carried out on the night of 
September 30, 2007, 
with the 0.84-m telescope.  A $1024 \times 1024$ CCD camera was used with
a plate scale of $0.43''$/pixel. About 4 h of data were obtained
using a Johnson $B$ filter.

\begin{figure}[!t]\centering
\includegraphics[width=8.cm]{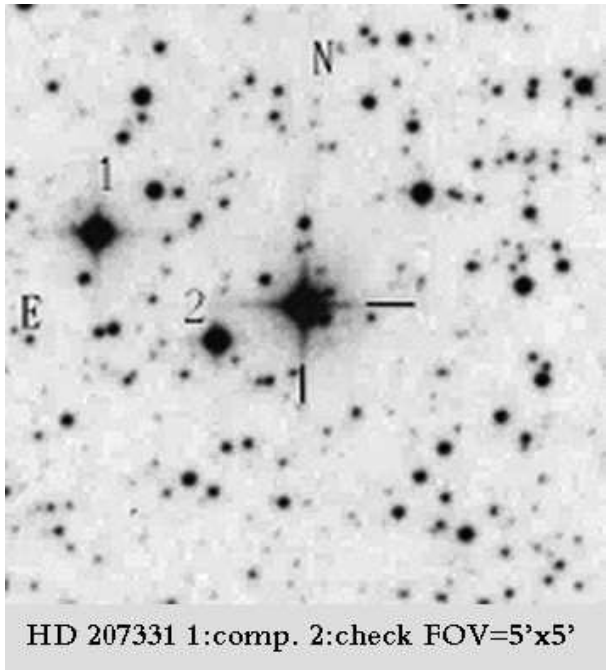} \caption{The
finding chart of HD 207331 (the central star). 
1 stands for the comparison star, and 2 for the check star. Some properties
of the stars are listed in Table 1. North is up and East is left.}
\end{figure}

\bigskip
Fig.~1 shows the finding chart of HD 207331.  The coordinates, $V$
magnitudes and identifications of the stars marked as 1 and 2  in
Fig.~1  are listed in Table 1. The acquired images were reduced in
the standard way using the IRAF package. Aperture photometry was
applied to extract the instrumental magnitudes of the stars. The
differential magnitudes were normalized by subtracting the mean of
differential magnitudes for the night.

\bigskip
 The differential light curve HD 207331 - Comparison is illustrated in Fig.~2.
 An exposure time of about 15 s was 
 applied (these data
 are depicted with dots in Fig.~2). The 3 min binned data are shown by asterisks.
 As can be seen, the oscillations of HD 207331 are clearly
inferred. The magnitude differences between the comparison and check
stars were also derived in order to confirm their constancy. No
indications of photometric variability for these stars were found.

\begin{figure}[!t]
\begin{center}
\includegraphics[width=8.0cm]{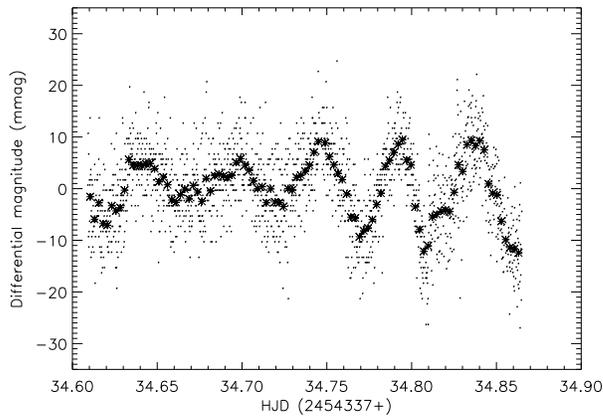}
\caption{CCD differential light curve HD 207331 - Comparison  }
\end{center}
\end{figure}

\subsection*{$uvby$ differential photometry}

The star was also observed during three nights in 
November 2007. The observations were 
performed with the 1.5-m telescope and the
six-channel Str\"omgren spectrophotometer (Schuster \& Nissen 1988).
The observing routine consisted of five 10 s integrations of the
star from which five times one 10 s integration of sky was
subtracted. Two constant comparison stars were observed as well. The
star was monitored for about 3.5 h on November 11, for about 4 h on
November 18, and 3.5 h on November 19. A set of standard stars was
also observed each night to transform instrumental observations onto
the standard system. Figure 3 shows the differential light curves in
the $y$ filter of HD 207331 for the three nights of our
observations. The comparison star C1 is HD 208310 ($=$BD$+$44 3980,
$V=8.43$, spectral type A0), while the comparison star C2 corresponds to HD
209113 ($=$BD$+$44 4012, $V=8.42$, spectral type A2). The light curves HD 207331~-~C1 and HD 207331~-~C2 are shown in the three top and three middle
plots of Fig.~3 respectively, with appropriate correction for
atmospheric extinction.  A multiperiodic characteristic of the star
with at least two beating periods can be inferred from these light
curves. The CCD differential photometry already showed such
behaviour for the light curve of HD 207331 (see Fig. 2).

\medskip
 The magnitude differences between the comparison stars, C1 - C2,
were also derived in order to confirm their constancy. As can be
seen in Figure 3 (the three bottom plots), no indications of
photometric variability for these stars were found.

\begin{figure}[!t]
\begin{center}
\includegraphics[width=8.0cm]{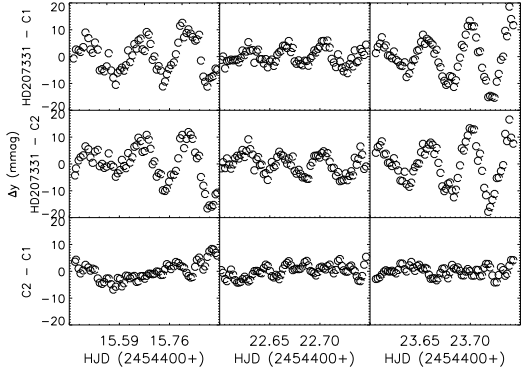}
\caption{The $y$ differential light curves with respect to the
reference stars C1 and C2 (top and middle plots, respectively). The
bottom plots are for C1 - C2.}
\end{center}
\end{figure}

\subsection{Absolute Str\"omgren photometry}

Preliminary standard photometry of HD207331, taken on the same
nights as the differential photometry, is the following (in mag):  ($V$,
$b$--$y$, $m_1$, $c_1$) = (8.329, 0.125, 0.150, 1.018) with the
standard errors of a single observation being ($\pm0.005$, 0.002,
0.002, 0.008) for 214 observations over the three independent
nights.

\medskip
To deredden this photometry, the $\beta$ index is not very useful
since HD207331 falls in the spectral range A0-A3, near the maximum
of the hydrogen-line absorption, where this index is not very
sensitive, changing from 
merely a temperature indicator for the
latter stellar types to 
merely a luminosity indicator for the O- and
B-type stars (Crawford 1978, 1979). So, a first estimate for the
reddening, $E(B$--$V)(l, b)_\infty$, has been taken from the
reddening maps of Schlegel et al. (1998) via the web-page calculator
of the NED (NASA/IPAC Extragalactic Database).  This reddening is
then reduced by a factor $\lbrace1-exp[-d \sin |b|/H]\rbrace$, where
{$b$} and {$d$} are the Galactic latitude and distance,
respectively, assuming that the Galactic dust layer has a scale
height $H = 125$ pc. According to SIMBAD $b = -7.823\,^\circ$ for
HD207331, and for a first estimate of this star's distance the
Hipparcos parallax is used to give $d = 302$ pc.  Then,
$E(B$--$V)(l, b)_\infty = 0.495$ mag, and so $E(B$--$V) = 0.139$
mag. However, Arce \& Goodman (1999) caution that the Schlegel et
al.~(1998) reddening maps overestimate the reddening values when the
color excess $E(B$--$V)$ is more than $\approx 0.15$ mag. Hence,
according to Schuster et al.~(2004), a slight revision of the
reddening estimate has been adopted via an equation, $E(B$--$V)_{\rm
A} = 0.10 + 0.65(E(B$--$V)-0.10)$ when $E(B$--$V) > 0.10$, otherwise
$E(B$--$V)_{\rm A} = E(B$--$V)$, where $E(B$--$V)_{\rm A}$ indicates
the adopted reddening estimate.  This leads to $E(B$--$V)_{\rm A} =
0.125$ mag for HD207331, and $E(b$--$y) = 0.093$ mag, from the
relation $E(B$--$V) = 1.35 E(b$--$y)$ of Crawford~(1975).  Then,
according to the relations:  $V_0 = V - 4.3 E(b$--$y)$, $m_0 = m_1 +
0.3 E(b$--$y)$, and $c_0 = c_1 - 0.2 E(b$--$y)$ (Str\"omgren 1966,
Crawford 1975), a first estimate for the dereddened photometry of
HD207331 is obtained:  ($V_0$, $(b$--$y)_0$, $m_0$, $c_0$) = (7.930,
0.032, 0.178, 0.999).  From Table~II of Crawford (1978) this leads
to $\beta \approx 2.852$, and from his Table~V, $M_V \approx 1.044$
mag, assuming that HD207331 is of luminosity class V, leading to an
improved distance of $d = 238$ pc. This process has then been
iterated twice more to a consistent solution:  $E(B$--$V)_{\rm A} =
0.110$ mag, $E(b$--$y) = 0.081$ mag, ($V_0$, $(b$--$y)_0$, $m_0$,
$c_0$) = (7.980, 0.044, 0.174, 1.002), $\beta = 2.854$, $M_V =
1.058$ mag, and $d = 242$ pc.  These final intrinsic colors of
HD207331 are very consistent with the spectral type of A0 given by
SIMBAD, according to Table~II of Crawford (1978).

\medskip
At the effective temperature of an A0-type star, the line blanketing
due to metallic atomic absorption lines is negligible, and so the
$m_1$ index of the Str\"omgren system is no longer useful for
providing stellar metallicity measures.  So, for HD207331 a solar
metallicity has been assumed for the following analyses.

\begin{table}[!t]
\begin{center}
\caption{Detected frequencies in the 
current study.  S/N is the
signal-to-noise ratio in amplitude. Var. is the percent of the total
variance explained by each peak. {\it $\nu_{a}$} is a possible
oscillation frequency.}
\begin{tabular}{lrcccc}
\hline
   &Freq. & A  & $\varphi$&S/N&Var. \\
 &($\mu$Hz)& (mmag)& (rad)&&\% \\
\hline
$\nu_{1}$  &244.7 & 6.0&-1.2&4.5&52\\
$\nu_{2}$  &311.1 & 2.9&0.9&3.7&20\\
\it{\small $\nu_{ a}$}  &{\small \it 250.0}& {\small \it 2.5} &{\small \it 0.1}&{\small \it 2.0}&{\small \it 17}\\
\hline
\end{tabular}
\end{center}
\label{tab:dom}
\end{table}

\section{Frequency analysis}
The amplitude spectra of the differential time series were obtained
by means of an iterative sine wave fit (ISWF; Ponman 1981) and  the
software package Period04 (Lenz \& Breger 2005). In both cases, the
frequency peaks are obtained by applying a non-linear fit to the
data. Both procedures allow to fit all the frequencies
simultaneously in the magnitude domain. Since both packages yielded
similar results, we present the spectral analysis only in terms of
ISWF. The amplitude spectrum
 of the differential light curve HD 207331 -
C1 is shown in the top panel of Figure 4.
 The subsequent panels in the
figure, from top to bottom, illustrate the prewhitening process of
the frequency peaks in each amplitude spectrum. The same procedure
was followed as explained in Alvarez et al. (1998). In particular,
according to these authors, a peak is considered as  significant when
the signal-to-noise ratio in amplitude is larger than 3.7. The
oscillation frequencies detected in HD 207331 are listed in Table 2.
S/N is the signal-to-noise ratio  after the prewhitening process.
Also shown is the fraction of the total variance contributed by each
peak.

\medskip
 The highest amplitude peak
is located at 244.7 $\mu$Hz (21.1 cd$^{-1}$), and the next
significant frequency is located at 311.1 $\mu$Hz (26.9 cd$^{-1}$) .
We note, however, that the second frequency is at the limit of our
detection level. Given our poor window function, with only three
nights of observations, it is not possible to detect more
frequencies in HD 207331.  Nevertheless, in the residual amplitude
spectrum, after prewhitening  $\nu_{2}$, another peak at {\it
$\nu_{a}=250$} $\mu$Hz seems to be present but with only a $S/N$ of
2.0.

\begin{figure}[!t]
\begin{center}
\includegraphics[width=9.1cm]{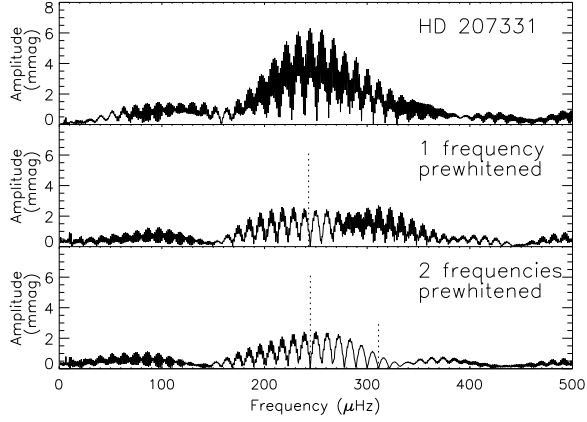}
\caption{Amplitude spectrum  of HD 207331 and the prewhitening
process of the detected peaks. }
\end{center}
\end{figure}

\section{Conclusions}
We have presented an analysis of  CCD  observations and $uvby$
differential photometry of the new $\delta$ Scuti star HD 207331,
which has been found to be a multiperiodic pulsator with at least
two modes of oscillations. A first estimate for the interstellar
reddening has been taken from the reddening maps of Schlegel et al.
(1998) via the web-page calculator of the NED database. The
resulting interstellar excess is $E(b-y)=0.081$ mag, and the
dereddened photometry is consistent with the classification of an early A star.

\bigskip
To date, our observations represent the most extensive work on HD~207331.
Beyond this,  more observations, better distributed in time,
are needed for an improved understanding of this interesting object.

\acknowledgments{ We would like to thank the assistance of the staff
of the OAN-SPM during the observations. This paper was partially
supported by Papiit IN108106,  and by CONACyT project 49434-F. We
thank the anonymous referee for his valuable comments which helped
us to improve the manuscript.}

\References{
Alvarez, M., Hern\'andez, M. M., Michel, E., et al. 1998, A\&A ,340, 149\\
Arce, H. G., \& Goodman, A. A. 1999, ApJ, 512, L135\\
Crawford, D. L. 1975, PASP, 87, 481\\
Crawford, D. L. 1978, AJ, 83, 48\\
Crawford, D. L. 1979, AJ, 84, 1858\\
Lenz, P., \& Breger, M. 2005, CoAst, 146, 53\\
Perryman, M. A. C., Lindegren, L., Kovalevsky, J.,  et al. 1997, A\&A, 323, L49\\
Ponman, T. 1981, MNRAS, 196, 583\\
Schlegel, D. J., Finkbeiner, D. P., \& Davis, M. 1998, ApJ, 500, 525\\
Schuster, W. J., Beers, T. C., Michel, R., Nissen, P. E., \&
Garc\'{\i}a, G. 2004, A\&A, 422, 527\\
Schuster, W. J., \& Nissen, P. E. 1988, A\&AS, 73, 225\\
Str\"omgren, B. 1966, Ann. Rev. Astron. Astrophys., 4, 433 }

\end{document}

%% file: CoAst_regularlayo.tex
\renewcommand{\chaptermark}[1]{\markboth{#1}{}}

\newcommand{\kopf}{\small\itshape Comm. in Asteroseismology \\ Vol. number, publication date (will be inserted in the production process)}

\newcommand{\Authors}[1]{\begin{center}\normalsize\bf\sf #1 \end{center}}

\renewcommand{\author}[1]{\begin{center}\normalsize\bf\sf #1 \end{center}}
\newcommand{\Address}[1]{\begin{center}\small\sf #1 \end{center}}

\newcommand{\Objects}[1]{{\vspace{2mm}\small \noindent  \hspace*{0mm}Individual Objects: } \small #1 \normalsize}

\renewenvironment{abstract}{\section*{Abstract}\normalsize\sf}{}
\newcommand{\References}[1]{\begin{flushleft}{\large References\\}\vspace*{2mm}\small #1 \end{flushleft}}

\newcommand{\chapterCoAst}[2]{\chapter[\sf\normalsize #1\\ \footnotesize \hspace*{5mm}by #2 \sf\normalsize][]{#1\\}\rhead[\fancyplain{}{\sf\footnotesize \center{#1}}]{\fancyplain{}{\sffamily\thepage}}\lhead[\fancyplain{\kopf}{\sffamily\thepage}]{\fancyplain{\kopf}{\sf\footnotesize \center{#2}}}}

\renewcommand{\chaptername}{}

\newcommand{\acknowledgments}[1]{\vspace*{5mm}\noindent  \textbf{Acknowledgments.} #1}